\documentclass[notitlepage,showpacs,aps,prl,onecolumn,showkeys,nofootinbib,nobibnotes]{revtex4-1}
\usepackage[dvipdfmx]{graphicx}
\usepackage{url}

\usepackage{lineno}
\modulolinenumbers[5]
\newcounter{bla}

\usepackage{graphicx,color}
\usepackage{amssymb,amsfonts,amsmath}
\usepackage[dvipdfmx]{hyperref}
\title{}
\begin{document}

\title{Designed-walk replica-exchange method  for simulations of complex systems \\
}
% \tnotetext[mytitlenote]{Fully documented templates are available in the elsarticle package on \href{http://www.ctan.org/tex-archive/macros/latex/contrib/elsarticle}{CTAN}.}

\author{ Ryo Urano$^{\rm 1}$ and Yuko Okamoto$^{{\rm 1,2,3,4}}$ } 
\affiliation{$^1$Department of Physics, Graduate School of Science, Nagoya University, Nagoya, Aichi 464-8602, Japan}    
 \affiliation{$^{\rm 2}$Structural Biology Research Center, Graduate School of Science, Nagoya University, Nagoya, Aichi 464-8602, Japan} 
\affiliation{$^{\rm 3}$Center for Computational Science, Graduate School of Engineering, Nagoya University, Nagoya, Aichi 464-8603, Japan}
\affiliation{$^{\rm 4}$Information Technology Center, Nagoya University, Nagoya, Aichi 464-8601, Japan
}


% \fntext[myfootnote]{Since 1880.}


\begin{abstract}
We propose a new implementation of the replica-exchange method (REM) in which replicas follow a  pre-planned route in temperature space instead of a random walk.
Our method satisfies the detailed balance condition in the proposed route.
The method forces tunneling events between the highest and lowest temperatures to happen with an almost constant period.
The number of tunneling counts is proportional to that of the random-walk REM multiplied by the square root of moving distance in temperature space. 
We applied this new implementation to two kinds of REM and compared the results with those of the conventional random-walk REM.
The test system was a two-dimensional Ising model, and our new method reproduced the results of the conventional  random-walk REM and could adjust  the tunneling counts by two times or more than that of the random-walk REM by replica-exchange attempt frequency. 
\end{abstract}


% \linenumbers

\maketitle

\section*{Introduction}
\label{sec-1}
The enhancement of configurational sampling ensures accuracy and efficiency of molecular dynamics (MD) or Monte Carlo (MC) simulations. Replica-exchange method (REM)\cite{rem1,swendsen1986replica,rem3,rem4} (or parallel tempering) is a popular way to improve efficiency of configurational sampling (for reviews, see, e.g., Refs.\cite{mitsutake_generalizedensemble_2001,rem_iba2001extended}).
 This method achieves a random walk in temperature space and allows the system to overcome energy barriers between local-minimum free energy states. 
The number of tunneling events\cite{berg1992multicanonicaltunnel,mitsutake2003replicatunnel} is defined to be the number of times where the simulation visits from the lowest temperature through the highest temperature and back to the lowest temperature (hereafter, we refer to this number as tunneling counts). 
The tunneling count is a measure of configurational sampling efficiency: the more tunneling counts are obtained with a fixed number of simulation steps, the better the sampling is.
However, as the number of replicas increases, tunneling events require more steps by random walk.
This is a difficulty for large systems because they require many replicas in general. 

    In the Metropolis criterion with pseudo random numbers, randomness prevents the design of a trajectory in temperature space. 
Moreover, the moving distance by a random walk is proportional to the square root of the number of trials. 
Therefore, previous improvements of tunneling count and efficiency in the REM was fulfilled by approaches such as the temperature selection\cite{trebst2004optimizing,katzgraber2006feedback,trebst2006optimized,nadler2007dynamics}, attempts of non-neighboring pairs, and increase of acceptance probabilities. 
An example is all-paring method, which tries all combinations of temperature pairs\cite{calvo2005all,brenner2007accelerating}. 
Another example is Gibbs sampling heat-bath replica-exchange method\cite{chodera2011replica}. 
Other examples are to employ the global balance condition such as replica permutation method\cite{itoh2012replica} with Suwa-Todo algorithm\cite{suwa2010markov} and all possible pair exchange\cite{kondo2013enhanced}.
These methods reported higher tunneling counts than the conventional REM.
However, a production of an equilibrium state without random walk leads to another possibility of an improved REM, which enables a  simulation to determine a trajectory of replicas in a temperature space. 
In fact, some of special manipulation of state changes without the detailed balance condition shows a rapid convergence for thermal equilibrium\cite{bernard2009event,turitsyn2011irreversible}. Based on Chaotic Boltzmann machine\cite{suzuki2013chaotic,suzuki2013monte}, we have also proposed the deterministic replica-exchange method (DETREM)\cite{2014arXiv1412.6959U}, which performs  replica exchange based on a differential equation without pseudo random numbers. This method produced the same efficiency as the conventional REM.

Recently, Spill $et~ al$. showed that a planned route trip in only one randomly selected replica during a simulation showed an improvement compared with the conventional REM\cite{spill2013convective}. 
We can generalize this idea so that planned route for all replicas will realize the maximum efficiency with wide configurational sampling, 
and we here propose the designed-walk replica-exchange method (DEWREM) by even-odd sequential replica exchange. 
\section*{Methods}
\label{sec-2}
We now give the details of our methods.
We prepare $M$ non-interacting replicas at $M$ different
 temperatures.
Let the label $i$ (=1, $\cdots$, $M$) stand for the replica index
 and label $m$  (=1, $\cdots$, $M$) for the temperature index. 
% Here, $i$ and $m$ are related by the permutation functions by
% \begin{eqnarray}
% \label{fpermu}
% \begin{cases}
% i=i(m) \equiv f(m), \\
% m=m(i)  \equiv f^{-1} (i),  
% \end{cases}
% \end{eqnarray}
% where $f(m)$ is a permutation function of $m$ and $f^{-1}(i)$ is the inverse.
We represent the state of the entire system of $M$ replicas by $X = \left\{x_{m(1)}^{[1]}  , \cdots, x_{m(M)}^{[M]} \right\}$, where $x_m^{[i]} =\left\{q^{[i]}, p^{[i]}\right\}_m$ are the set of coordinates $q^{[i]}$ and momenta $p^{[i]}$ of particles in replica $i$ (at temperature $T_m$).
The probability weight factor for state $X$ is given by a product of Boltzmann factors: 
\begin{eqnarray}
W_{\rm REM}(X)&=\displaystyle \prod_{i=1}^M \exp{[-\beta_{m(i)} H(q^{[i]} , p^{[i]})]},
% &=\prod_{i=1}^M \exp{[-\beta_{m)} H(q^{[i(m)]} , p^{[i(m)]})]}, 
\end{eqnarray}
where $\beta_m (=1/k_{\rm B} T_m)$ is the inverse temperature and $H(q,p)$ is the Hamiltonian of the system. 
We consider exchanging a pair of replicas $i$ and $j$
 corresponding to temperatures $T_m$ and $T_n$, respectively: 
\begin{equation}
\label{lstate}
 X =  \left\{ \cdots, x_m^{[i]}  , \cdots,
 x_n^{[j]}, \cdots \right\} \rightarrow   X^\prime =  \left\{ \cdots, x_m^{[j]^\prime}  , \cdots,
 x_n^{[i]^\prime}, \cdots \right\}, 
\end{equation}
where $x_n^{[i]^\prime} \equiv \left\{q^{[i]}, p^{[i]^\prime}\right\}_n ,x_m^{[j]^\prime} \equiv \left\{q^{[j]}, p^{[j]^\prime}\right\}_m$, and $p^{[j]^\prime}=\sqrt{\frac{T_m}{T_n}} p^{[j]},p^{[i]^\prime}=\sqrt{\frac{T_n}{T_m}} p^{[i]}$ \cite{rem4}.

Here, the transition probability $\omega (X\rightarrow X^\prime)$ of Metropolis criterion for replica exchange is given by 
\begin{eqnarray}
\label{metro}
 \omega (X\rightarrow X^\prime )  = {\rm min}\left(1, \frac{W_{\rm REM} (X^\prime)}{W_{\rm REM} (X)}\right)  = {\rm min}(1,\exp(- \Delta  )) ,
\end{eqnarray}
 where 
\begin{eqnarray}
\label{ddelta}
\Delta =\Delta_{m,n}   = (\beta _n - \beta_m ) ( E(q^{[i]}) - E(q^{[j]})   ). 
\end{eqnarray}
Because each replica visits various temperatures followed by the transition
 probability of Metropolis algorithm, REM performs a random walk in temperature space.

We now review two REMs, which are based on random walks in temperature space.
Without loss of generality, we can assume that $M$ is an even integer and that 
$T_1 < T_2<\cdots <T_M$.
The conventional REM\cite{rem1,swendsen1986replica,rem3,rem4} is performed by repeating the following two steps:

\begin{enumerate}
\item We perform a conventional MD or MC simulation of replica $i~ (=1,\cdots, M)$ at temperature $T_m~ (m=1,\cdots,M)$ simultaneously and independently for short steps.
\item Pairs of exchange attempts are selected in replica pairs with neighboring temperatures, for example, for the odd pairs ($T_1,T_2$), ($T_3,T_4$),$\cdots$, ($T_{M-1},T_M$) or even pairs ($T_2,T_3$), ($T_4,T_5$),$\cdots$, ($T_{M-2},T_{M-1}$).
\end{enumerate}
All the replica pairs thus selected are attempted to be exchanged according to the Metropolis transition probability in Eqs. (\ref{metro}) and (\ref{ddelta}) with $n=m+1$. 

We repeat Steps 1 and 2 until the end of the simulation. The canonical ensemble at any temperature is reconstructed by reweighting techniques\cite{ferrenberg_optimized_1989,kumar1992weighted,wham3}.

We next present the deterministic replica-exchange method (DETREM)\cite{2014arXiv1412.6959U}. Only Step 2 is different from the conventional REM.
At first, we introduce an internal state $y_{m,n}$ as an index of a pair of replicas $i$ and $j$ at temperatures $T_m$ and $T_n$, and consider the following differential equation: 
\begin{equation}
\label{timeinte}
\frac{dy_{m,m+1} }{dt}= \sigma _{m}\frac{1}{1+ {\rm exp}(\Delta_{m,m+1}  )}, 
\end{equation}
 where $t$ is a virtual time, $\Delta_{m,m+1}$ is the same as in Eq. (\ref{ddelta}) with $n=m+1$, and 
the signature $\sigma_{m} $ of the pair of $(T_m,T_{m+1})$  changes to 1 or $-1$ to control the signature of the change of $y_{m}$ which monotonically increases or decreases. 
In Step 2, instead of applying the Metropolis criterion in Eqs. (\ref{metro}) and (\ref{ddelta}), 
we solve the differential equation in Eq. (\ref{timeinte}) for the
 internal states $y_{m,m+1} \in \left\{-1,1\right\}$  for  $(T_{m},T_{m+1})$, where the total number of internal
 states is $M$-1 with the following pairs: (1,2), (2,3), $\cdots$, ($M$-1,$M$) for the random-walk DETREM  and the pairs: (1,2), (3,4), $\cdots$, ($M$-1,$M$) and (2,3), (4,5), $\cdots$, ($M$-2,$M$-1) for designed-walk REM.
The replica exchange is done as follows\cite{2014arXiv1412.6959U}:
\begin{eqnarray*}
\displaystyle 
 {\rm if\ updated}\ y_{m,m+1} \gtrless  \pm 1, &\  {\rm then}   &\  {(T_{m}, T_{m+1} )  \rightarrow (T_{m+1}, T_{m} )}, \\  & & \ y_{m,m+1} \leftarrow y_{m,m+1} \mp1, \sigma _m \leftarrow \mp 1  .
%& \rm updated&\ y_i > 1 \Longrightarrow   {(T_{(i)}, T_{(i+1)} ) \rightarrow (T_{(i+1)}, T_{(i)} )}, \ y_i \leftarrow y_i -1, \sigma _i \leftarrow -1  .
% & \rm updated&\  y_i < -1 \Longrightarrow  {(T_{(i)}, T_{(i+1)} ) \rightarrow (T_{(i+1)}, T_{(i)} )}, \ y_i \leftarrow y_i +1, \sigma _i \leftarrow +1 
\end{eqnarray*}
For the random-walk DETREM, if $y_{m,m+1}$ performs exchanges, $y_{m+1,m+2}$ is not time evolved and $y_{m+2,m+3}$ is evolved to avoid the leap exchange of temperature such as from $T_m$ to $T_{m+2}$.

Finally, the designed temperature walk can be implemented to both conventional REM and DETREM (and other REMs) as follows.
Namely, the designed-walk replica-exchange method (DEWREM) is performed by repeating the following steps.

1. We perform a conventional MD or MC simulation of replica $i~ (=1,\cdots, M)$ at temperature $T_m~ (m=1,\cdots,M)$ simultaneously and independently for short steps.

2.  Replica exchange is attempted for all the odd pairs ($T_1$,$T_2$), ($T_3$,$T_4$), $\cdots$, ($T_{M-1}$,$T_M$). 

3. Repeat Steps 1 and 2 until all odd pairs perform replica exchange exactly once. Namely, once a pair is exchanged, the exchanged pair stops exchange attempts and keep performing the simulation in Step 1 with the new temperatures.  
Replica exchange attempt in Step 2 is repeated until all the other odd pairs finish exchanges. 

4--6. Repeat Steps 1--3 where the odd pairs in Steps 2 and 3 are now replaced by the even pairs ($T_2$,$T_3$), ($T_4$,$T_5$), $\cdots$, ($T_{M-2}$,$T_{M-1}$).

7. The cycle of Steps 1 to 6 is repeated until the number of cycles is $M$, which is equal to the tunneling count and all replicas have the initial temperatures.

8.  Begin the above cycle of Steps 1--7 with Steps 1 to 3 and Steps 4 to 6 interchanged. \\
These eight steps are repeated until the end of the simulation.

\begin{figure}[htb]
\centering
\includegraphics[height=0.3\textheight]{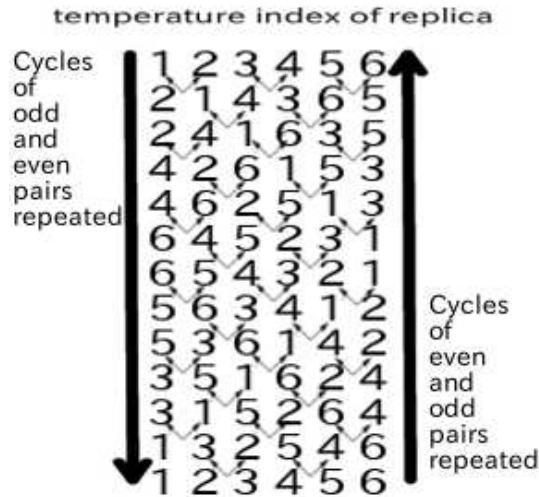}
\caption{\label{seq}An schematic picture of  time series of temperature indices in DEWREM with 6 replicas. The left cycle begins with the temperature exchange of odd index pairs $(T_1, T_2),~(T_3,T_4)$, and $(T_5,T_6)$, then tries with even pairs $(T_2, T_3)$ and $(T_4,T_5)$. The right cycle begins with even pairs and next tries odd pairs. They are the reverse cycles of each other and their combination satisfies the detailed balance condition of replica exchange.}
\end{figure}

The schematic picture of this procedure is shown in Fig. \ref{seq}.
We remark that Step 8, namely,  reversing the cycle of Steps 1--3 and 4--6, is necessary for the detailed balance condition, because the entering states are the same as leaving states. 
For example, the state $(x_1^{[1]}, x_2^{[2]}, x_3^{[3]}, x_4^{[4}], x_5^{[5]}, x_6^{[6]})$ is reached from only two states $(x_2^{[1]}, x_1^{[2]}, x_4^{[3]}, x_3^{[4]}, x_6^{[5]}, x_5^{[6]})$, $(x_1^{[1]}, x_3^{[2]}, x_2^{[3]}, x_5^{[4]}, x_4^{[5]}, x_6^{[6]})$ only and makes transition to the two states as shown in Fig. \ref{seq}, where $x_m^{[i]}$ is the state of replica $i$ at temperature $T_m$.
This exchange procedure satisfies the detailed balance condition for replica and temperature pair because the trials of exchange pair 
\begin{eqnarray}
\gamma &\Bigl( & i(m) \rightarrow  i(m+1) \Bigr) \omega\left({(x^{i(m)}_{m}, x^{i(m+1)}_{m+1} ) \rightarrow (x^{i(m)}_{m+1}, x^{i(m+1)}_{m} )} \right) \nonumber \\
&=& \gamma \Big(i(m+1) \rightarrow i(m) \Bigr) \omega\left(   (x^{i(m)}_{m+1}, x^{i(m+1)}_{m} )  \rightarrow (x^{i(m)}_{m}, x^{i(m+1)}_{m+1} ) \right) 
\end{eqnarray}
is equal in the route as is shown in Fig. \ref{seq}, where $\gamma (i(m) \rightarrow i(m+1))$ is the selected probability of the exchange attempt.
Note that for exactly the same conditional probability of odd pair replica exchange, ($T_1$,$T_2$), ($T_3$,$T_4$), $\cdots$, ($T_{M-1}$,$T_M$), $\Delta$ is  $\sum_{k=1}^{M/2} (\beta _{2k-1} - \beta_{2k})( E_{2k}-E_{2k-1} )$, where  $k=1, \cdots, M/2$. 
However, because each replica is non-interacting with other replicas, waiting for other exchanges does not influence the transitions of others.
% In other words, since neighboring pair is independent among exchange pairs, we use the individual replica exchange instead of the same conditional probability with the waiting for other exchanges. 
% This means that because the important point of the internal time integration is the ratio of states corresponding to the Boltzmann distribution, this individual pair exchange during the above sequential exchange is sufficient for the achievement of equilibrium. 

This sequential exchange achieves one tunneling count when $M$ cycles  for each replica are finished.
In theory, the estimated ratio of tunneling count between the odd-even sequential exchange and the conventional random walk is given by
\begin{equation}
\displaystyle \frac{TC_{\rm sequential} }{ TC_{\rm random\ walk}}=\frac{\displaystyle \frac{N_{{\rm trial}}\times P^{{\rm DEW}}_{{\rm correction}}}{2M} }{\displaystyle 
\frac{\sqrt{N_{{\rm trial}}}\times P^{{\rm RW}}_{{\rm correction}}}{2M}} \propto \sqrt{N_{{\rm trial}}},
\end{equation}
where $N_{{\rm trial}}$ is the number of exchange attempts, $P^{{\rm DEW}}_{{\rm correction}}$ is the correction for waiting for all the replica exchanges in Steps 3 and 6, and $P^{{\rm RW}}_{{\rm correction}}$ is the correction for the deviation of random-walk probability  from the value 1/2.
\section*{Results}
\label{sec-3}
In order to test the effectiveness of the present methods, we applied them to the 2-dimensional Ising model. The lattice size in a square lattice was 128 (hence, the number of spins  was $N=128^2=16384$). 
We have performed conventional random-walk simulation and designed-walk (DEWREM) simulation of both conventional REM and DETREM.
We have also performed a mixed random-walk and designed-walk simulation of DETREM, where we repeated the two walks alternately.
The total number of replicas $M$ was 40 and the temperatures were 1.50, 1.55, 1.60, 1.65, 1.70, 1.75, 1.80, 1.85, 1.90, 1.94, 1.98,  2.01, 2.04, 2.07, 2.10, 2.13, 2.16, 2.19, 2.22, 2.25, 2.28, 2.31, 2.34,  2.358, 2.368, 2.38, 2.40, 2.42, 2.44, 2.47, 2.51,  2.57, 2.63 ,2.69, 2.75, 2.82 ,2.90,  3.00, 3.10, and 3.15. Boltzmann constant $k_{\rm B}$ and coupling constant $J$ were set to 1. Thus, $\beta = 1/ k_{\rm B} T = 1/T =\beta ^*$, and the potential energy is given by
$E({\bf s})= - \sum_{<i,j>} s_i s_j$, 
where $s_i=\pm 1$, and the summation is taken over all the nearest-neighbor  pairs in the square lattice.

For the conventional random-walk REM and DETREM, replica-exchange attempt was made every 1 MC step. 
One MC step here consists of one Metropolis update of spins.
The total number of MC steps for all the simulations was 100,000,000.
To integrate Eq. (\ref{timeinte}), we used the fourth-order Runge-Kutta method with virtual time step $dt=1$. 
For DEWREM, replica-exchange attempt was made every 10, 20, 50, 100, and 150 MC steps in the conventional REM simulations and every 10, 20, 50, and 100 MC steps in the DETREM simulations (see Table \ref{tc}). 
The mixed-walk simulation was performed in which after $4M(=160)$ even-odd or odd-even cycles of designed-walk simulations (replica-exchange attempt was made at every 20 MC steps) were performed, 200,000 MC steps (which roughly corresponds to $2M$ cycles)  of random-walk simulations (replica-exchange attempt was made at every 1 MC step) were performed, and then this procedure was repeated.
For reweighting analyses\cite{ferrenberg_optimized_1989,kumar1992weighted,wham3,shirts2008statistically}, the total of 10,000 spin state data were taken with a fixed interval of 1,000 MC steps at each temperature from the REM simulations.

\begin{figure}[htb]
\centering
\includegraphics[width=0.6\textheight]{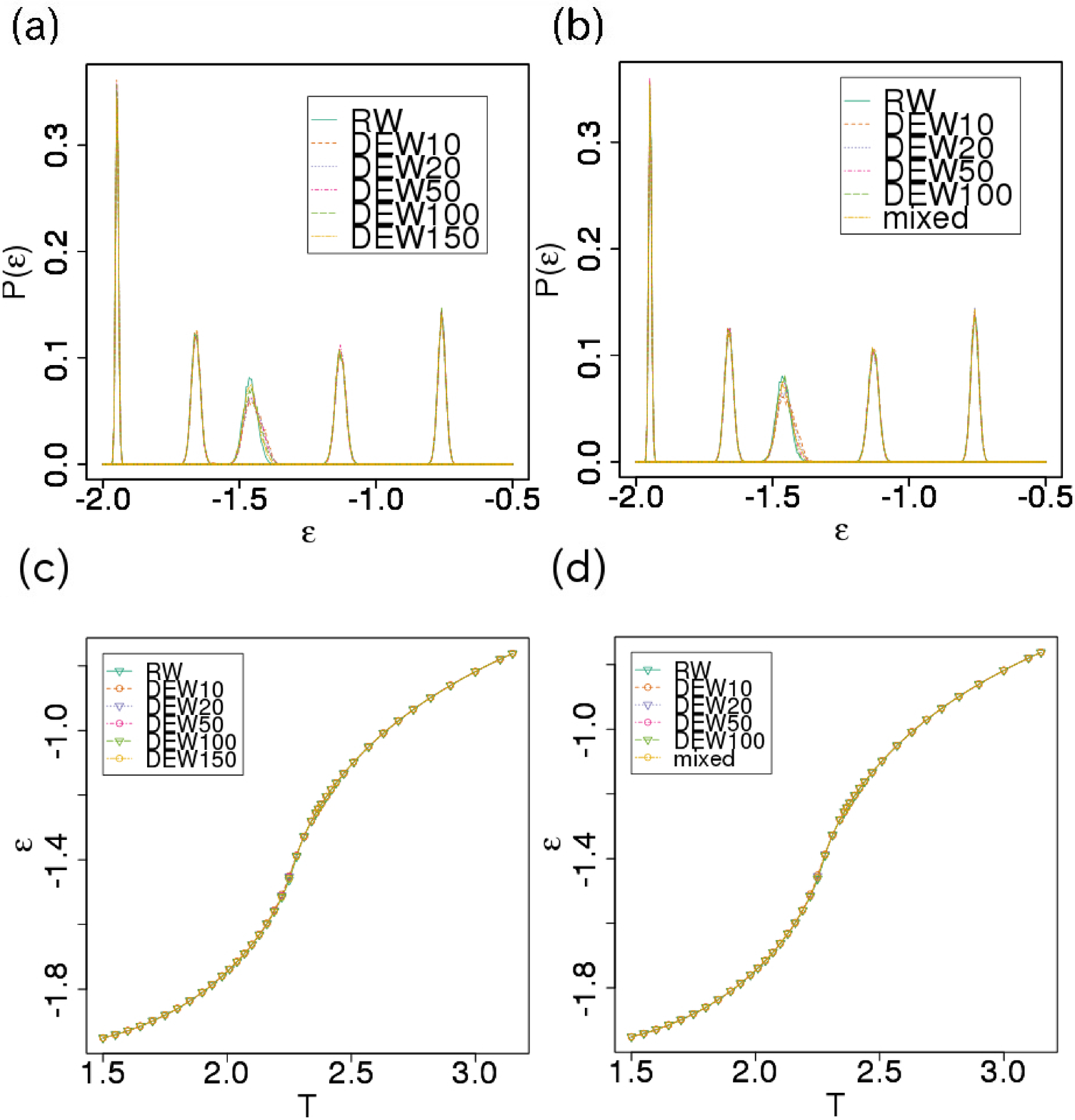}
\caption{\label{enecapa}Probability distributions of energy density at four temperatures (from left to right, 1.50, 2.10, 2.25, 2.47, and 3.15) obtained from (a) REM and (b) DETREM simulations including the mixed-walk simulation, and average energy density as a function of $T$ from (c) REM and (d) DETREM simulations including the mixed-walk simulation and the error bars are smaller than the symbols.}
\end{figure}

\begin{figure}[htb]
\centering
\includegraphics[width=1.0\textwidth]{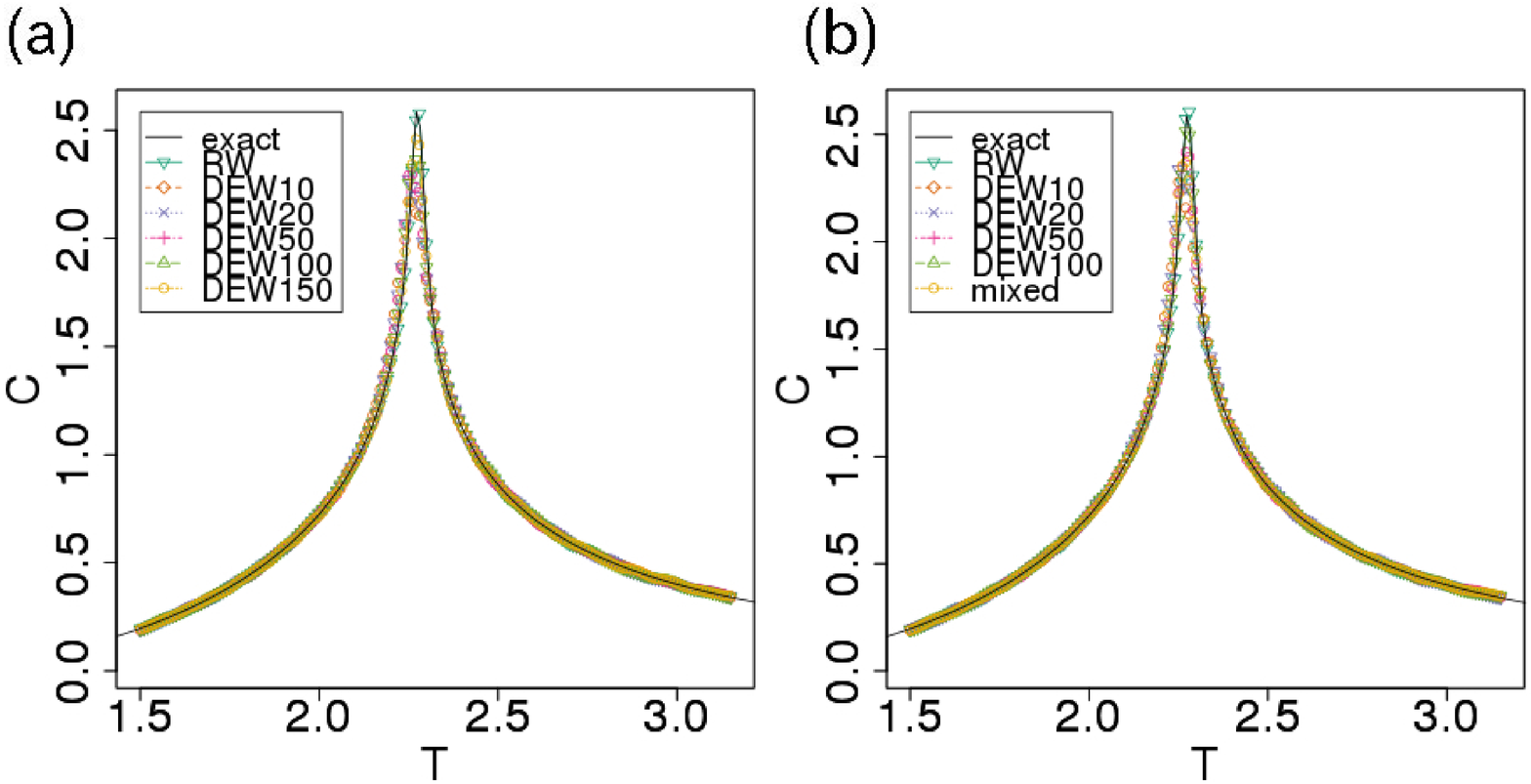}
\caption{\label{capa}Specific heat $C$ as a function of $T$ from the  (a) REM, (b) DETREM simulations including the mixed-walk simulation. The error bars are smaller than the symbols. The exact results for $L$ =128 (black curves) were obtained by Berg's program \cite{berg2004markov} based on Ref. \cite{ferdinand1969bounded}.}
\end{figure}

Table \ref{tc} lists the mean tunneling counts in temperature space and energy space per replica for each method, which is the number of times where the replicas visit from the lowest temperature through the highest temperature and back to the lowest, and the mean energy at the lowest temperature through the mean energy at the highest temperature and back to the mean energy at the lowest temperature during the simulation, respectively. 
The mean tunneling counts in temperature space per replica of the designed-walk simulations at every 100 MC attempts were about four times larger than random-walk Metropolis REM and twice larger than random-walk DETEM. 
On the other hand, the mean tunneling counts in energy space per replica of the designed-walk simulations at every 100 MC attempts were about six times larger than random-walk Metropolis REM and twice larger than random-walk DETEM. 
These large tunneling counts imply that in designed-walk  method all replicas traversed more efficiently in temperature space, and our design to maximize the tunneling counts for all replicas without random walks was successful. For the mixed-walk simulation, the maximum tunneling count was about twice larger than that of random-walk DETREM. The mean tunneling count was almost the same as that of designed-walk DETREM. 
We next examine physical quantities obtained from the designed-walk simulations with various replica-exchange attempt frequencies and mixed walk simulations and compare them to those from the conventional random-walk simulations.
Fig. \ref{enecapa}(a) and Fig. \ref{enecapa}(b) compare the canonical probability distributions of energy density $\epsilon=E/N$ at four temperatures as a function of $T$ and the average total energy density $\epsilon$ as a function of $T$  obtained from the random-walk and designed-walk simulations of REM and DETREM, for DETREM including the mixed-walk simulation.
Most of the probability distributions of energy density are the same.
However, in the DEWREM simulations, the probability distributions with high frequency replica-exchange attempts at $T=2.25$ near the exact critical temperature $T_c=2.269$ are deformed slightly compared to the results of the random-walk simulation.
This can also confirm the results of heat capacity as shown in Fig. \ref{capa}.
  Fig. \ref{enecapa}(c) and Fig. \ref{enecapa}(d) show the average total energy density $\epsilon=E/N$ as a function of $T$ from the same simulations.
They were obtained by the reweighting techniques\cite{ferrenberg_optimized_1989,kumar1992weighted,wham3,shirts2008statistically}.
The average total energy densities in all the simulations are the same. 

 Fig. \ref{capa}(a) and Fig. \ref{capa}(b) show the specific heat $C$ as a function of $T$ during the conventional REM simulations  and the DETREM simulations, respectively.
This shows that designed-simulation with shorter replica-exchange interval such as every 10 and 20 MC steps underestimated the heat capacity near the critical temperature although the transition point is sufficiently similar to the exact critical temperature at $T_c=2.269$. 
As the intervals of replica-exchange attempts are longer, the accuracy of heat capacity is higher.
Moreover, the combination of the random-walk and designed walk also increased the accuracy.
This suggests that the designed-walk replica-exchange attempts caused correlation between replicas. The correlation seems to be very strong near the critical temperature. As a result, the heat capacity is underestimated slightly.
However, these physical quantities show that mixed-walk simulation can increase the accuracy of results and the number of tunneling counts and DEWREM simulation is suited for simulations with longer replica-exchange time intervals between replica-exchange attempts. 
\section*{Conclusions}
\label{sec-4}
In this article, we proposed an algorithm by a designed temperature route for replica-exchange methods.
The aim for developing this method was to maximize the tunneling counts.
We reproduced the results of conventional REM. 
Our designed-walk simulations showed that the tunneling counts increased by at least twice more than the random-walk simulations.
This method also showed that in REM replicas have no interaction among replicas but replica-exchange trials with artificial orders cause a correlation between replicas.
By decreasing the correlations introducing mixed-walk simulation,  using longer- interval replica-exchange trials, or other paring of replicas, this method will be more efficient.

In a future work, we will give the formulation for designed routes for multidimensional replica-exchange method\cite{sugita2000multidimensional}, which is a power tool in massive parallel computing\cite{rhee2003multiplexed,vogel2013generic,vogel2014scalable}.
This will be useful for creating another design route for non-neighboring update for exchanges.

\begin{table}[!tbp]
\caption{The mean number of tunneling counts per replica.\label{tc}} 
\begin{center}
\begin{tabular}{l|p{2pt}||p{2pt}|p{2pt}||p{2pt}|p{2pt}|p{2pt}|p{2pt}|p{2pt}|p{2pt}|p{2pt}|p{2pt}|p{2pt}}
\hline\hline
 \multicolumn{1}{l||}{TC}&\multicolumn{4}{c||}{Random walk}&\multicolumn{7}{c||}{Designed walk}&\multicolumn{1}{c}{Mixed}\\
 \hline 
 \multicolumn{1}{l||}{T\_TC}&\multicolumn{2}{|c|}{Met}&\multicolumn{2}{|c||}{DETREM}&\multicolumn{4}{|c|}{Met}&\multicolumn{3}{|c||}{DETREM}&\multicolumn{1}{|c}{DETREM}\tabularnewline
    \hline 
   \multicolumn{1}{l||}{Interval}&\multicolumn{1}{|c|}{1}&\multicolumn{1}{|c||}{100}&\multicolumn{1}{|c||}{1}&\multicolumn{1}{|c||}{100}
   &\multicolumn{1}{|c|}{20} &\multicolumn{1}{|c|}{50} &\multicolumn{1}{|c|}{100}&\multicolumn{1}{|c|}{150}
   &\multicolumn{1}{|c|}{20} &\multicolumn{1}{|c|}{50} &\multicolumn{1}{|c||}{100}
   &\multicolumn{1}{|c}{1 $\&$ 20}\tabularnewline
   \hline 
  \multicolumn{1}{l||}{Mean}&\multicolumn{1}{|c|}{173}&\multicolumn{1}{|c||}{37}&\multicolumn{1}{|c||}{178}&\multicolumn{1}{|c||}{58}
  &\multicolumn{1}{|c|}{292}&\multicolumn{1}{|c|}{197}&\multicolumn{1}{|c|}{131}&\multicolumn{1}{|c|}{99}
  &\multicolumn{1}{|c|}{231} &\multicolumn{1}{|c|}{144} &\multicolumn{1}{|c||}{93}
  &\multicolumn{1}{|c}{293}\tabularnewline
 \multicolumn{1}{l||}{$\pm$ SD }&\multicolumn{1}{|c|}{ 10}&\multicolumn{1}{|c||}{3}&\multicolumn{1}{|c||}{9}&\multicolumn{1}{|c||}{5}
 &\multicolumn{1}{|c|}{56}&\multicolumn{1}{|c|}{41}&\multicolumn{1}{|c|}{27}&\multicolumn{1}{|c|}{21}
 &\multicolumn{1}{|c|}{ 48} &\multicolumn{1}{|c|}{ 30} &\multicolumn{1}{|c||}{ 20} 
 &\multicolumn{1}{c}{6}\tabularnewline 
 \hline
 \multicolumn{1}{l||}{E\_TC}&\multicolumn{2}{|c|}{Met}&\multicolumn{2}{|c||}{DETREM}&\multicolumn{4}{|c|}{Met}&\multicolumn{3}{|c||}{DETREM}&\multicolumn{1}{|c}{DETREM}\tabularnewline
    \hline 
   \multicolumn{1}{l||}{Interval}&\multicolumn{1}{|c|}{1}&\multicolumn{1}{|c||}{100}&\multicolumn{1}{|c||}{1}&\multicolumn{1}{|c||}{100}
   &\multicolumn{1}{|c|}{20} &\multicolumn{1}{|c|}{50} &\multicolumn{1}{|c|}{100}&\multicolumn{1}{|c|}{150}
   &\multicolumn{1}{|c|}{20} &\multicolumn{1}{|c|}{50} &\multicolumn{1}{|c||}{100}
   &\multicolumn{1}{|c}{1 $\&$ 20}\tabularnewline
   \hline 
  \multicolumn{1}{l||}{Mean}&\multicolumn{1}{|c|}{55}&\multicolumn{1}{|c||}{13}&\multicolumn{1}{|c||}{55}&\multicolumn{1}{|c||}{34}
  &\multicolumn{1}{|c|}{80}&\multicolumn{1}{|c|}{79}&\multicolumn{1}{|c|}{77}&\multicolumn{1}{|c|}{70}
  &\multicolumn{1}{|c|}{81} &\multicolumn{1}{|c|}{79} &\multicolumn{1}{|c||}{69}
  &\multicolumn{1}{|c}{75}\tabularnewline
 \multicolumn{1}{l||}{$\pm$ SD }&\multicolumn{1}{|c|}{ 5}&\multicolumn{1}{|c||}{2}&\multicolumn{1}{|c||}{5}&\multicolumn{1}{|c||}{4}
 &\multicolumn{1}{|c|}{8}&\multicolumn{1}{|c|}{6}&\multicolumn{1}{|c|}{7}&\multicolumn{1}{|c|}{5}
 &\multicolumn{1}{|c|}{ 6} &\multicolumn{1}{|c|}{ 6} &\multicolumn{1}{|c||}{ 4} 
 &\multicolumn{1}{c}{5}\tabularnewline 
 \hline
 \hline
\end{tabular}
\end{center}
\begin{flushleft}
T\_TC, E\_TC, Interval, SD, and Met stand for tunneling counts in temperature space, tunneling count in energy space, the number of MC steps between replica-exchange attempts, standard deviation, and REM based on Metropolis criterion, respectively. The frequency (1 $\&$ 20) of Mixed means that it was 1 MC step for random walk REM and 20 MC steps for designed-walk REM. 
\end{flushleft}
\end{table}

\section*{Acknowledgments}
\label{sec-5}
Some of the computations were performed on the supercomputers at the
Institute for Molecular Science, at the Supercomputer Center, Institute
for Solid State Physics, University of Tokyo, and Center for
Computational Sciences, University of Tsukuba.
This work was supported, in part, Grants-in-Aid for Scientific Research (A) (No. 25247071),
for Scientific Research on Innovative Areas (\lq\lq Dynamical Ordering \& Integrated
Functions\rq\rq ), Program for Leading Graduate Schools \lq\lq Integrative Graduate Education and Research in Green Natural Sciences\rq\rq, and for the Computational Materials Science Initiative, and for High Performance Computing Infrastructure from the Ministry of
Education, Culture, Sports, Science and Technology (MEXT), Japan.

\bibliographystyle{apsrev4-1}
\bibliography{ref}

\end{document}